# Low Complexity Sphere Decoding for Spatial Multiplexing MIMO

Vadim Neder, Doron Ezri and Motti Haridim

1. Abstract

In this paper we present a novel method for decoding multiple input - multiple output (MIMO) transmission, which combines sphere decoding (SD) and zero forcing (ZF) techniques to provide near optimal low complexity and high performance constant time modified sphere decoding algorithm. This algorithm was designed especially for large number of transmit antennas, and allows efficient implementation in hardware. We do this by limiting the number of overall SD iterations. Moreover, we make sure that matrices with high condition number are more likely to undergo SD.

2. Introduction

One of the most promising MIMO transmission methods is spatial multiplexing (SM). In SM the transmitter endowed with $M$ transmit antennas, transmits $M$ independent information stream, one from each antenna. In the case of SM, the receiver endowed with $N \geq M$ receive antennas, is to decode the transmitted information streams. It is known that the optimal solution to the decoding of SM signals is maximum-likelihood (ML), which involves exhaustive search in multiple dimensions.

Sphere decoding *[1]* is an iterative method for the computation of the ML estimator in SM MIMO. However, one of the severe problems in the implementation of SD lies in the fact that the number of iterations per realization is neither defined nor bounded. Thus, usually, SD methods are not suitable for hardware implementation.

V. Neder and D. Ezri are with Runcom Technologies LTD, {vadimne,dorone}@runcom.co.il.
M. Haridim is with the Department of Communication Engineering at Holon Institute of Technology, mharidim@hit.ac.il.

A few resent results pertaining to fixed complexity SD are the following. In [6] the K-Best lattice decoder with breadth-first tree search was presented. This method uses the breadth-first tree search technique which introduces fixed throughput. In this method the best K candidates, which have the smallest overall Euclidian distance, are kept at each search level, therefore a fixed amount of nodes are visited each time. The disadvantage of this is that its K parameter cannot be defined analytically and it is also very dependant of the channel condition.

In [7] the authors propose the depth-first tree search SD. This is a straightforward way of enforcing a run-time constraint is to terminate the search, on a symbol vector by symbol vector basis, after a maximum number of visited nodes. The detector then returns the best solution found so far, i.e., the current ML and counter-hypotheses. This method also, as in [6] can degrade detection performance in case of bad channel condition.

In [8] the unconstrained list sphere detector with a search method that is bounded independently per search level is proposed. The bound is determined based on the distribution of the candidates found in each search level for the large number of detected sub-carriers. It is shown that the search process cannot be bounded for the first search level without a substantial performance loss. This method exploits the main idea of [7] but with lower upper bound, also it doesn't provide the constant rate, but only bounds it.

In [9] it is shown that diversity achieving scheme may be devised, by combination of the low complexity zero forcing (ZF) algorithm and ML detection. This method is based of division of the channel matrices into 2 sets according to the condition number. Matrices with condition number lower than a predefined threshold are ZF decoded, while the others are ML decoded. However, this result does not allow hardware implementation of SD for the ML estimates since again, the number of iterations is not defined. Moreover, threshold based technique implies receiver calibration, which should be recalibrated for different channel conditions.

## 3. Review of Prominent Spatial Multiplexing Decoding Algorithms

The mathematical model for the received vector $\mathbf{y}$ in the case of SM is

$$\mathbf{y} = \mathbf{Hs} + \rho \mathbf{v} \qquad (1)$$

where $\mathbf{H}$ is the channel matrix, $\mathbf{s}$ is the transmitted signal vector and $\rho \mathbf{v}$ vector of independent complex valued Gaussian random variables (RV) each with variance $\rho^2$. A schematic of the SM scenario is given in Figure 1.

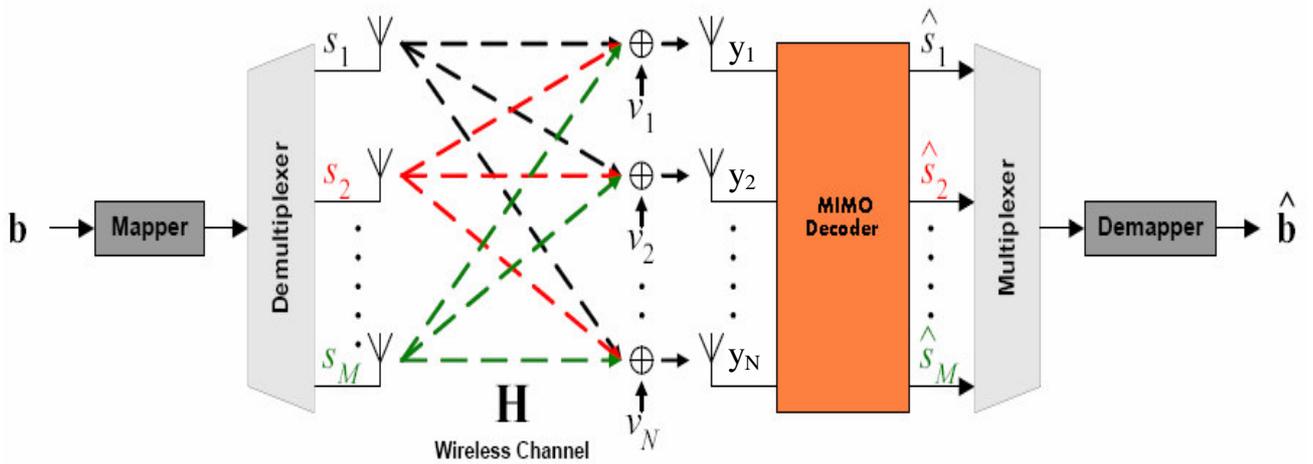

*Figure 1: Schematic representation of an SM system.*

### 3.1. Zero Forcing

The linear zero forcing (ZF) computes the least square estimator [3]

$$\hat{\mathbf{s}}_{ZF} = \mathbf{H}^{+} \mathbf{y} \qquad (2)$$

where $\mathbf{H}^{+}$ denotes the left pseudo-inverse of $\mathbf{H}$. The estimator $\hat{\mathbf{s}}_{ZF}$ then undergoes standard processing as in the single input single output case (SISO). The complexity of finding the ZF estimate is essentially determined by the complexity of finding the pseudo-inverse of the matrix

**H** in (1). For large matrices, the simplest way of calculating the pseudo-inverse is by means of QR factorization, **H** = **QR**. It can also be calculated in a more stable way (which avoids inverting the upper triangular matrix **R**) by means of the singular value decomposition (SVD) of **H**. The ZF algorithm is not optimal in the case of MIMO, but remains attractive due to its low implementation complexity. The problem with the ZF approach is evident when the channel matrix **H** is ill conditioned (small determinant), corresponding to strong correlation between the channels. In this case, the entries of $\mathbf{H}^+$ in (2) are large. This leads to large noise at the output of the ZF estimator. The ZF solution provides diversity order of $N-M+1$ and array gain of $\frac{N-M+1}{M}$.

### 3.2. Maximum Likelihood

We now address the optimal ML decoder for SM. In this case, the optimal log likelihood ratio (LLR) of a bit $b$ in the data-stream **s** is defined by

$$LLR(b) = \log \frac{\Pr\{b=1 \mid \mathbf{y}\}}{\Pr\{b=0 \mid \mathbf{y}\}}. \tag{3}$$

Applying Bayes formula we obtain

$$LLR(b) = \frac{\Pr\{b=1 \mid \mathbf{y}\}}{\Pr\{b=0 \mid \mathbf{y}\}} = \frac{\frac{\Pr\{\mathbf{y} \mid b=1\} \cdot \Pr\{b=1\}}{p\{\mathbf{y}\}}}{\frac{\Pr\{\mathbf{y} \mid b=0\} \cdot \Pr\{b=0\}}{p\{\mathbf{y}\}}} = \frac{\sum_{\mathbf{s}:b=1} \Pr\{\mathbf{y} \mid \mathbf{s}\}}{\sum_{\mathbf{s}:b=0} \Pr\{\mathbf{y} \mid \mathbf{s}\}}. \tag{4}$$

Using (1) and (5) we obtain

$$LLR(b) = \frac{\sum_{\mathbf{s}:b=1} e^{-\frac{\|\mathbf{y}-\mathbf{Hs}\|^2}{\rho^2}}}{\sum_{\mathbf{s}:b=0} e^{-\frac{\|\mathbf{y}-\mathbf{Hs}\|^2}{\rho^2}}} \tag{5}$$

which can be rewritten using max-log approximation as

$$LLR(b) = \log \frac{e^{-\min_{s:b=1}\left(\frac{\|\mathbf{y}-\mathbf{Hs}\|^2}{\rho^2}\right)}}{e^{-\min_{s:b=0}\left(\frac{\|\mathbf{y}-\mathbf{Hs}\|^2}{\rho^2}\right)}} = \frac{1}{\rho^2}\left(-\min_{s:b=1}\|\mathbf{y}-\mathbf{Hs}\|^2 + \min_{s:b=0}\|\mathbf{y}-\mathbf{Hs}\|^2\right). \tag{6}$$

However, this approach quickly becomes impractical when the number of streams or number of constellation points is large as it requires exhaustive search.

The ML solution provides diversity order of N and array gain of $\frac{N}{M}$.

### 3.3. Sphere Decoding

SD is an iterative method that converges to the ML when the number of iterations is not bounded. In SD, the multidimensional search implies by the ML criterion is transformed to multiple searches in one complex dimension.

The building block of the optimal LLR is the search for the minimizer of the functional

$$\min_{\mathbf{s}\in\Gamma}\|\mathbf{y}-\mathbf{Hs}\|^2 \tag{7}$$

over some set of points (2-dimensional QAM) $\Gamma$.

Denoting the ZF solution as $\hat{\mathbf{s}}$, the cost functional in (7) may be rewritten as

$$\|\mathbf{H}(\hat{\mathbf{s}}-\mathbf{s})\|^2 = (\hat{\mathbf{s}}-\mathbf{s})^*\mathbf{H}^*\mathbf{H}(\hat{\mathbf{s}}-\mathbf{s}) \tag{8}$$

Note that since $\mathbf{H}^*\mathbf{H}$ is a positive definite symmetric matrix, it can always be decomposed to $\mathbf{U}^*\mathbf{U} = \mathbf{H}^*\mathbf{H}$ where $\mathbf{U}$ is an upper triangular matrix with real diagonal (this can be done by applying the QR decomposition on $\mathbf{H}$). Thus, the cost functional (8) turns to

$$(\hat{\mathbf{s}} - \mathbf{s})^* \mathbf{U}^* \mathbf{U} (\hat{\mathbf{s}} - \mathbf{s}) \qquad (9)$$

The special structure of $\mathbf{U}$ allows writing (9) explicitly, for the $N \times 2$ case, as

$$\sum_{i=1}^{2} u_{ii}^2 \left| s_i - \hat{s}_i + \sum_{j=i+1}^{2} \frac{u_{ij}}{u_{ii}} (s_j - \hat{s}_j) \right|^2 = u_{22}^2 |s_2 - \hat{s}_2|^2 + u_{11}^2 \left| s_1 - \hat{s}_1 + \frac{u_{12}}{u_{11}} (s_2 - \hat{s}_2) \right|^2 \qquad (10)$$

We begin with searching for points $\mathbf{s}$ for which the cost functional (10) is smaller than an arbitrary $r^2$. Taking only the first term in the sum (10) we obtain a necessary (but not sufficient) for a point $\mathbf{s}$ to have a cost smaller than $r^2$ as

$$u_{22}^2 |s_2 - \hat{s}_2|^2 < r^2 \Rightarrow |s_2 - \hat{s}_2|^2 < \frac{r^2}{u_{22}^2} \qquad (11)$$

which implies that a necessary condition is that $s_2$ lies within a circle about the ZF solution $\hat{s}$. If there are no points in the set $\Gamma$ satisfying (11), the magnitude of $r$ is increased and the algorithm starts all over. In the case there are points that satisfy the condition, we peak one of them and use it to produce a similar condition on $s_1$ (for the specific $s_2$ chosen) through (10) as

$$u_{22}^2 |s_2 - \hat{s}_2|^2 + u_{11}^2 \left| s_1 - \hat{s}_1 + \frac{u_{12}}{u_{11}} (s_2 - \hat{s}_2) \right|^2 < r^2 \qquad (12)$$

which implies that $s_1$ should lie within a circle about ZF that depends on the $s_2$ chosen. If there are no points $s_1$ satisfying the condition, we turn to the next point $s_2$ satisfying (11), Otherwise we have a point $\mathbf{s}$ with cost smaller than $r_2$, dubbed candidate. We compute the cost of this point say $\tilde{r}^2 < r^2$ and repeat the algorithm with $\tilde{r}^2$.

Eventually, $r^2$ will be small enough such that no points with smaller cost exist and the minimizer is the candidate of the last iteration. Surely, if no points exist for a certain $r^2$ and no candidates have been found in previous iterations, $r^2$ should be increased.

One of the major problems with the SD algorithm is that the number of iterations is not constant and may significantly vary between matrices. This makes hardware implementation of SD very difficult. When the number of iterations of SD algorithm is not limited, the array gain and diversity order are the same as for the ML.

## 4. The Combined SD ZF Method

The proposed method assumes constant hardware clock budget for the decoding of the $K$ matrices, each with dimensions $N \times M$. We further assume that the clock budget is larger than that needed for ZF decoding of all matrices.

The method is based on several ideas. The first is that matrices with high condition number should be likely to undergo SD. The second is that the hardware clock budget must remain constant for the decoding of $K$ matrices. Note that we do not attempt to construct an SD algorithm with finite number of iterations for each matrix, but restrict the number of overall iterations for the decoding of multiple matrices.

Following this line of thought, the proposed method sums up to the following steps:

1. Compute the linear ZF decoder

$$\hat{\mathbf{s}} = \mathbf{H}^{+}\mathbf{y} \qquad (13)$$

for each of the $K$ SM inputs (or matrices). We note an SVD based approach is preferred here since it expedites the calculation of the condition number.

2. Order the $K$ matrices according to the condition number, in descending order (largest first). This way the more problematic matrices in terms of decoding are first in order.

3. Apply SD to the matrices according to the above-mentioned order until the hardware clock budget runs out. We note that the SD algorithm requires the ZF solution already obtained in the first step, so no waste of clock budget is done in the first step. Figure 2 shows possible HW implementation of the algorithm.

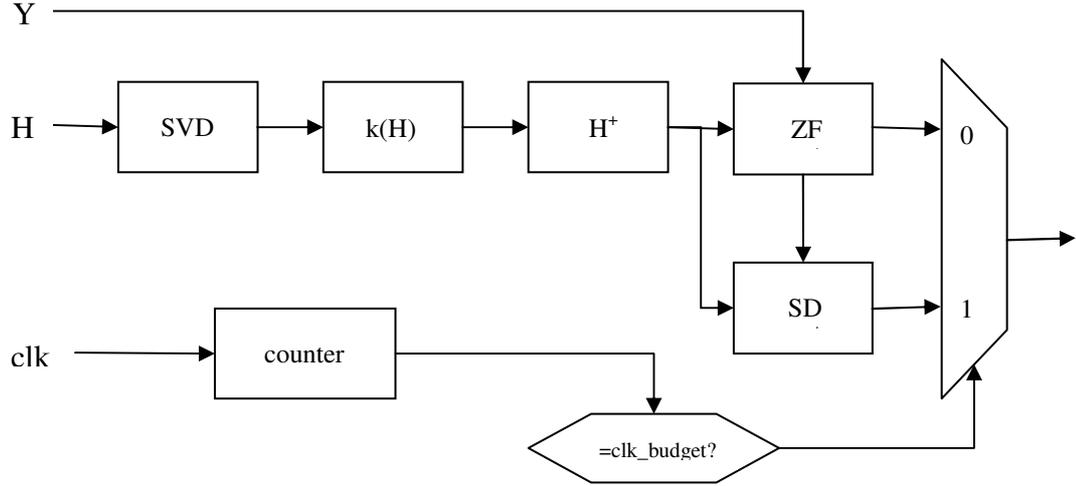

*Figure 2: Possible HW implementation of the algorithm.*

Thus, in the proposed algorithm, the matrices with high condition number are first to invoke the SD mechanism, which means efficient use of the hardware resources. The performance of the proposed algorithm is low bounded by that of ZF (in case the clock budget is identical to that required for ZF decoding), and high bounded by the performance of ML (in case the clock budget is sufficient for SD of all matrices).

We conclude with the understanding that the performance of the algorithm in actual scenarios is determined by the clock budget allocated and the distribution of the condition number of the channel matrices.

## 5. Simulation Results

Simulation results for the proposed algorithm are given in Figure 3. We define the parameter $n$ as

$$n = \frac{n_{TOT}}{n_{ZF}} \qquad (14)$$

where

$n_{TOT}$ - is the number of overall HW clocks, reserved for the decoding

$n_{ZF}$ - is the number of HW clocks, reserved for ZF decoding

The BER curves for ZF and ML are added to the figure for means of comparison. The figure shows the BER curves corresponding to the performance of the proposed algorithm with different clock budgets. Obviously the BER is smaller as the clock budget is increased. The most important result of our method we can see in Fig. 2, which shows, that when the matrices are sorted by their conditional number, a small fraction of them undergoing SD, but still we can get significant enhancement in the performance. Note further that in the case of *n*=10, where the average of 18% of the matrices is undergoing SD, the performance is almost identical to the optimal ML decoder.

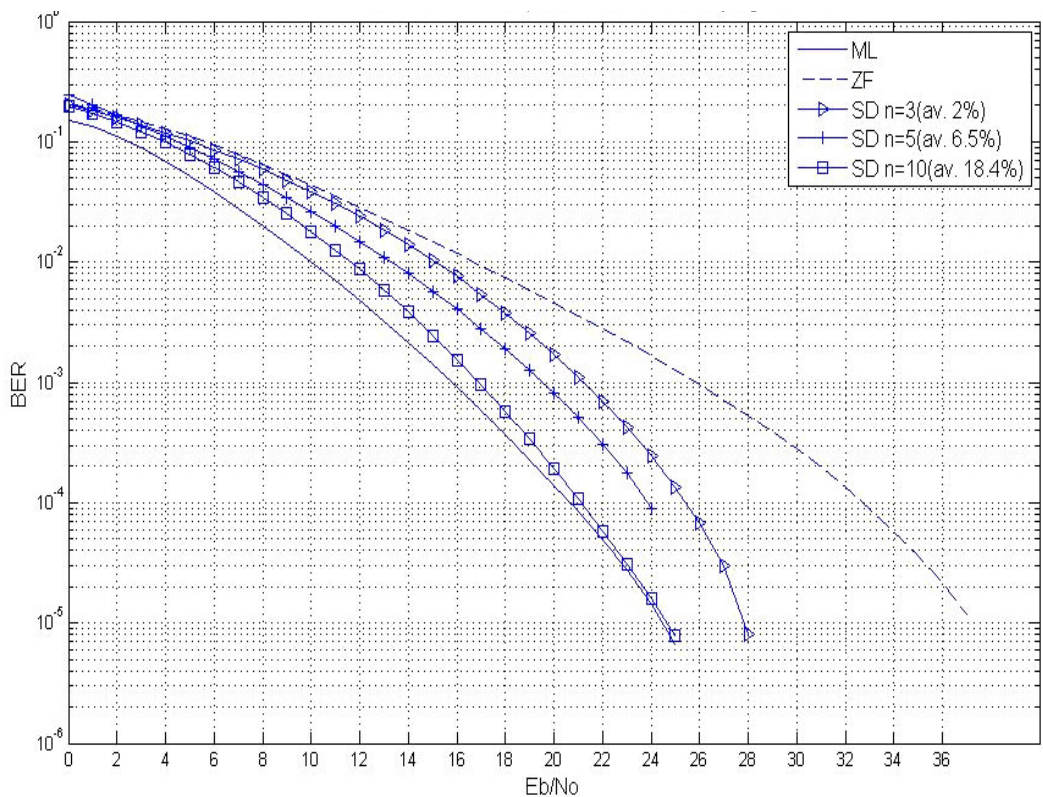

*Figure 3 Simulation results of ML, ZF and proposed SD algorithms in the Raleigh fading environment*

## 6. Discussion and Conclusions

In this paper we proposed a novel constant time modified SD algorithm for decoding MIMO transmission, which is upper bounded by exact ML solution, depending of overall number of iterations, reserved for the decoding. Higher number of overall iterations causes the algorithm to be closer to the optimal ML solution.

The future wireless communication systems are more likely to incorporate a large number of transmit and receive antennas. IEEE802.16 and WiMax standards are already discussing future user terminals and base stations with large number of antennas arrays. This kind of setup altogether with high rate QAM modulations schemes will make today's MIMO ML decoding algorithms a very problematic issue for the future HW implementation. The proposed algorithm allows the employment of the SD to a small portion of the matrices, allowing the accommodation of large antenna arrays featuring a large number of spatial streams.

## 7. References

[1] U. Fincke, M. Pohst, "Improved methods for calculating vectors of short length in a lattice, including a complexity analysis". Mathematics of Computation, vol. 44, pp. 463–471, 1985

[2] Foschini, G.J.: "Layered space-time architecture for wireless communication in a fading environment when using multiple antennas', Bell Lab. Tech. J., 1996, 1, (2), pp. 41–59

[3] M. Grotschel, L. Lov´asz, A. Schriver, "Geometric Algorithms and Combinatorial Optimization", Springer Verlag, 2nd ed., 1993

[4] T. Kailath, H. Vikalo, B. Hassibi, "MIMO Receive Algorithms" in Space-Time Wireless Systems: From Array Processing to MIMO Communications, (editors H. Bolcskei, D. Gesbert, C. Papadias, and A. J. van der Veen), Cambridge University Press, 2005

[5] Arogyaswami Paulraj, Rohit Nabat, Dhananjay Gore, "Introduction to Space-Time Wireless Communications", Cambridge University Press, Cambridge, UK, 2003

[6] Kwan-wai Wong, Chi-Ying Tsui, R. S.-K. Cheng, Wai Ho Mow, "A VLSI architecture of a K-best lattice decoding algorithm for MIMO channels", ISCAS (3) 2002: 273-276

[7] D. Garrett, L. Davis, S. ten Brink, B. Hochwald, and G. Knagge, "Silicon complexity for maximum likelihood MIMO detection using spherical decoding", IEEE Journal of Solid-State Circuits, vol. 39, pp. 1544–1552,Sept. 2004.

[8] B. M. Hochwald and S. ten Brink, "Achieving near-capacity on a multiple-antenna channel", IEEE Trans. Commun., vol. 51, no. 3, pp. 389{399, Mar. 2003

[9] Maurer, J., Matz, G., and Seethaler, D., "Low complexity and full diversity MIMO detection based on condition number thresholding", ICASSP 2007